\documentclass[12pt]{article}
\usepackage{arxiv}
\usepackage{algorithm}
\usepackage{algpseudocode}
\usepackage{amsmath,amssymb, amsfonts,amsthm}
\usepackage{hyperref}
\usepackage{booktabs}
\usepackage{dsfont}
\usepackage{blkarray}
\usepackage{caption}
\usepackage{subcaption}
\usepackage{secdot}

\usepackage{tikz}
\usepackage[square,numbers]{natbib}

\newtheorem{definition}{Definition}

\title{Federated Privacy-preserving Collaborative Filtering for On-Device Next App Prediction}

\author{Albert Sayapin\thanks{Corresponding author (\url{analystalb@gmail.com}) } \\
Skoltech, Moscow, Russia \\
\And
Gleb Balitskiy \\
Skoltech, Moscow, Russia\\
\And
Daniel Bershatsky\\
Skoltech, Moscow, Russia\\
\AND
Aleksandr Katrutsa \\
Skoltech, Moscow, Russia\\
\And 
Evgeny Frolov\\
Skoltech, Moscow, Russia\\
\And
Alexey Frolov \\
Skoltech, Moscow, Russia\\
\And
Ivan Oseledets \\
Skoltech, Moscow, Russia\\
\And
Vitaliy Kharin\\
An Independent Expert\\
}
\date{}

\newcommand\ddfrac[2]{\frac{\displaystyle #1}{\displaystyle #2}}

\newcommand{\diag}[1]{\text{diag}\left( #1 \right)}

\newcommand{\fref}[1]{Figure~\ref{#1}}
\newcommand{\sref}[1]{Section~\ref{#1}}

\newcommand{\inR}[1]{\in\mathbb{R}^{#1}}

\newcommand{\matr}[1]{\mathbf{#1}}  
\newcommand{\vect}[1]{\mathbf{#1}}  
\newcommand{\ind}[1]{{\left({#1}\right)}}
\newcommand{\bmatr}[1]{\bar{\mathbf{#1}}} 

\begin{document}


\maketitle

\begin{abstract}
In this study, we propose a novel SeqMF model to solve the problem of predicting the next app launch during mobile device usage.
Although this problem can be represented as a classical collaborative filtering problem, it requires proper modification  since the data are sequential, the user feedback is distributed among devices and the transmission of users' data to aggregate common patterns must be protected against leakage.
According to such requirements, we modify the structure of the classical matrix factorization model and update the training procedure to sequential learning.
Since the data about user experience are distributed among devices, the federated learning setup is used to train the proposed sequential matrix factorization model.
One more ingredient of the proposed approach is a new privacy mechanism that guarantees the protection of the sent data from the users to the remote server.
To demonstrate the efficiency of the proposed model we use publicly available mobile user behavior data.
We compare our model with sequential rules and models based on the frequency of app launches. 
The comparison is conducted in static and dynamic environments.
The static environment evaluates how our model processes sequential data compared to competitors. 
Therefore, the standard  train-validation-test evaluation procedure is used.
The dynamic environment emulates the real-world scenario, where users generate new data by running apps on devices, and evaluates our model in this case.
Our experiments show that the proposed model provides comparable quality with other methods in the static environment.
However, more importantly, our method achieves a better privacy-utility trade-off than competitors in the dynamic environment, which provides more accurate simulations of real-world usage.
\end{abstract}



\section{Introduction}
Today smartphones collect a lot of personal data. 
Based on the collected data, mobile OS can predict the next app that will be run by the user.
This internal service helps to manage available resources on the device better.
For example, mobile OS can offload some apps from memory and pre-load new ones in the background to save computational resources or prolong battery life.
Such gains are important to improve the user experience of using the device.

This study considers the next app prediction problem and proposes a novel approach to solve it.
The proposed approach is based on two ingredients.
The first part is a new SeqMF model that extends the existing federated matrix factorization approach~\cite{Ammad_2019_FedCF, Minto_2021_FedCF} to sequential learning setup.
And the second ingredient is a novel privacy mechanism to anonymize data that are sent from devices to the remote server. 
While it is possible to solve the next app prediction problem separately on each device and avoid transfer of user data to the remote server, previous studies in recommender systems~\cite{Su_CF_Survey_2009} suggest that users typically exhibit common patterns in their behavior. 
Therefore, utilizing collective behavioral data may help improve the quality of the predictive model.
This, in turn, requires information exchange with a remote server to enable collaborative filtering (CF) setup.

As a basis for our work, we use a special form of the federated learning setup~\cite{Abdulrahman_2021_Survey_FedLearn} adapted for the matrix factorization model. 
This idea was proved to be appropriate for solving general collaborative filtering problems~\cite {ammad2019fedMF} and was recently equipped with additional privacy protection~\cite{Minto_2021_FedCF}. 
The key distinction of our work is that we focus on the on-device next-app prediction problem that significantly differs from the general collaborative filtering problem.
This instructs using sequential data for model training and evaluation.
In addition, we revisit the problem of privacy-preserving federated learning and propose an improved QHarmony mechanism for the anonymization of transferred data.

Hence, \emph{the main contributions} of our work are twofold:
\begin{itemize}
\item We propose a computationally efficient collaborative filtering model SeqMF for the next app prediction problem. 
The model is based on a sequence-aware matrix factorization technique and is trained in a federated learning setup (\sref{sec:seqmf_next_app}).
\item  We equip the SeqMF model with a new Local Differential Privacy (LDP) algorithm based on the modified Harmony mechanism (\sref{sec::q_harmony}).
We demonstrate that it has a better privacy-utility trade-off compared to competitors (see \sref{subsec:results}).
\end{itemize}

\paragraph{Paper organization.}
\label{sec:problem_organization}
The structure of paper is organized as follows. 
We first introduce the related work on how recommender systems are used for the next app prediction and related privacy mechanisms, in \sref{Related_Work}. 
We then discuss the general problem statement and the feature of the next-app prediction problem in \sref{sec:problem_statement}. 
Then we elaborate on the new sequence-aware matrix factorization model and its federated learning for the next app prediction in~\sref{sec::background}. 
We introduce the novel $\epsilon$-LDP QHarmony privacy mechanism in \sref{sec::q_harmony}.
In \sref{sec:experiments} we describe the computational experiments and show the results in~\sref{subsec:results}. 
Finally, we summarize the obtained results in \sref{sec:conclusion}.

\section{Related Work}
\label{Related_Work}
Many algorithms for on-device app prediction were proposed in the last decade.
For example, \citet{Kamisaka} and \citet{Shin} adopted a naive Bayesian model with manually crafted features extracted from mobile log data which is not publicly available.
\citet{Zhang} built a Bayesian network relying on time, date, location, and previously used applications in inference. 
Several nearest neighbors-based methods were shown to provide great flexibility for the task~\cite{Liao, Ye, changmai2019device}.
\citet{Baeza-Yates} tested forecasting methods, including Tree Augmented Naive Bayes and a decision tree based on the C4.5 algorithm.
However, they were mostly designed to work locally on devices without data exchange with a remote server. 
Thus they can not enable the collaborative setup.  

The collaborative filtering formulations of the next app prediction problem were developed and studied in \cite{McAuley_2016_Fossil,rendle2010factorizing,Jannach_2018_Eval_SBRS}.
The federated learning paradigm adapted for the collaborative filtering task was recently explored by~\citet{ammad2019fedMF}. 
The authors proposed a special hybrid optimization scheme for a \emph{non-sequential} model that is trained in a federated learning setup and incorporates  personal data protection. 
Later, the data protection in this approach was further improved by~\citet{Minto_2021_FedCF} with a more sophisticated LDP mechanism.

In addition, many deep learning-based models were developed for the task of sequential recommendation. 
A comprehensive survey of such models is presented in~\citet{fang2020deep}. 
However, neither federated learning setup nor privacy-related concerns were explored in these works.
Thus, the applicability of deep learning models in real-world scenarios remains an open problem.

To protect sensitive user data during transfer in the collaborative filtering setup, LDP mechanism is used, since they do not assume the trusted third party (e.g server) in contrast to Centralized Differential Privacy (CDP) mechanism~\cite{DP_centr,DP_centr_book,DP_centr_2}.
Typically, LDP mechanism encrypts the sensitive user data on a device and only after that transfers them to the remote server.
Therefore, the encryption stage has to be  computationally and memory efficient.

The most famous deployment of LDP mechanism is RAPPOR algorithm~\cite{rappor}.
It is based on the Randomized Response mechanism~\cite{rr_mech}, which, later found its application in many LDP algorithms. 
However, this method is not suitable since it works with categorical data. 
For real-valued data, the Laplace mechanism~\cite{DP_centr_book} was adopted. 
This mechanism allows one to guarantee privacy but causes a strong drop in utility if the dimension of the transmitted data is large. 
Then, \citet{Duchi2016MinimaxOP} proposed improvement of the Randomized Response (RR) technique. 
In~\cite{Nguyen2016CollectingAA}, the authors showed that in some cases Duchi's mechanism does not preserve privacy. 
To address the revealed issues, they proposed a simple Harmony mechanism. 
Also, in \cite{Chen2020BreakingTC} it was suggested to use Kashin representation and quantization before applying RR. 
For now, asymptotically, this method achieves the best trade-off between utility, communication cost, and privacy budget. 
However, it has limitations for usage on devices due to high computational complexity.

\section{Problem statement}
\label{sec:problem_statement}
Let $M$ be the number of users and $N$ be the total number of apps available for installation.
From the users' history, we can compose the interaction binary matrix $\matr{A} = [a_{ij}]$ of size $M \times N$ such that $a_{ij} = 1$ if the $i$-th user has run the $j$-th app. 
Formally, assume we have a set of the installed apps indices $A_u$ on a user $u$ device such that $|A_u| \ll N$ and an \emph{ordered} set of the used apps $\mathcal{H}^\ind{t}_u = \{i_1, i_2, \ldots, i_p \}$ such that $i_k \in A_u$, which we call history of the used apps by the moment $t$.
Since matrix $\matr{A}$ is constructed from this historical data, it also depends on the parameter $t$ which we omit to simplify notation.

Our main task is to predict the next app that will be run on the particular user's device. 
Here, we see the sequential nature of the given data that is crucial for the high performance of the trained model.
Then we have to obtain a function that maps the available history or its part to the next app index $i^* \in A_u$.
Following previous studies~\cite{Koren_2008_ICF, ammad2019fedMF}, we construct 
the relevance score function~$r$ that depends on unknown embeddings of apps $\vect{q}_i, \; i=1,\ldots,N$ and of users $\vect{p}_u, \; u = 1,\ldots,M$.
The higher value of $r(\vect{q}_i, \vect{p}_u)$ is, the more relevant app $i$ to user $u$ is.
Therefore, the index of the predicted next app for current user $u$ is the index of the maximum element of $r(\vect{q}_i, \vect{p}_u)$ over all candidate apps $i \in A_u$.

However, we need to obtain the proper embedding vectors $\vect{p}_u$ and $\vect{q}_i$ from the observed history collected in matrix~$\matr{A}$.
The classical approach to finding these embeddings is to solve the following optimization problem~\cite{Koren_2008_ICF}:
\begin{equation}\label{eq:general_objective}
    \frac12 \sum_{u, i} c_{ui}(a_{ui} - r(\vect{q}_i, \vect{p}_u))^2 + \frac{\lambda}{2}\sum_{i }\| \vect{q}_i \|_2^2 + \frac{\lambda}{2} \sum_u\| \vect{p}_u \|_2^2 \to \min_{\{ \vect{q}_i, \vect{p}_u \}},
\end{equation}
where $c_{ui}$ measures a confidence in observing $a_{ij}$ and $r(\vect{q}_i, \vect{p}_u) = \vect{q}_i^\top \vect{p}_u$.

After computing optimal embeddings $\{ \vect{q}_i^*, \vect{p}_u^* \}$, we can suggest $n$ apps as candidates for a particular user as follows:
\begin{equation}
    \label{eq:prediction}
\operatorname{toprec}(u; n) = \arg\max_{i \in A_u}^n\left(r(\vect{q}^*_i, \vect{p}^*_u)\right),
\end{equation}
where $\arg\max\limits_{i \in A_u}^n\left(f(i)\right) = (i_1^*, i_2^*, \dots, i_n^*)$ is the ordered set of apps from $A_u$ such that if $f(i_k^*) \ge f(i_l^*)$, then $k < l$.

The standard method to solve problem~\eqref{eq:general_objective} is the alternating least squares (ALS) approach that updates $\vect{q}_i$ and $\vect{p}_u$~\cite{Koren_2008_ICF}.
However, since every user stores its own interaction history $\mathcal{H}_u^{(t)}$ locally on the device and all users have apps from a pre-defined set of available apps, it is natural to exploit the \emph{federated learning paradigm}~\cite{ammad2019fedMF, Minto_2021_FedCF} to solve~\eqref{eq:general_objective}.
In this paradigm, users' embeddings~$\vect{p}_u$ are updated locally on devices in a parallel asynchronous manner and apps' embeddings~$\vect{q}_i$ are updated globally in the remote server.
To perform such updates, we have to support the transmission of data from users' devices to the server and vice versa. Since sending data from a user to a third-party server can lead to personal data leaks, the proper privacy algorithm must be incorporated into the training procedure.
Moreover, the relevance score function $r$ has to be constructed to operate with sequential data corresponding to the history of interactions with apps.
Taking into account these requirements, we present a proper modification of~\eqref{eq:general_objective} in the next section.

\section{Federated learning of the sequence-aware matrix factorization model}
\label{sec::background}

In this section, we design a new objective that incorporates knowledge about users' sequential behavior into the framework of the objective~\eqref{eq:general_objective}.
After that, we present the details of the federated learning approach to get optimal users' and apps' embeddings. 
This approach combines the ALS step for updating users' embeddings locally and the gradient-based update of the apps' embeddings in the remote server.

\subsection{Sequence-Aware Matrix Factorization for the Next App Prediction}
\label{sec:seqmf_next_app}
 
\subsubsection{Objective function for sequential data}
\label{sec:seqmf}
Let $\matr{P} \in \mathbb{R}^{M \times d}$ and $\matr{Q} \in \mathbb{R}^{N\times d}$ be matrices composed row-wisely from users' and apps' embeddings, respectively. 
Since the users' history is distributed among the devices, we rewrite~\eqref{eq:general_objective} in a user-oriented form:
\begin{equation}\label{eq:gen_loss}
    \mathcal{L}(\matr{P}, \matr{Q}) = \frac{1}{2}\sum_u \left\|\vect{r}(\matr{Q}, \vect{p}_u)-\vect{a}_u\right\|^2_{\matr{C}_u} + \frac{\lambda}{2}\,\left(\sum_{u} \|\vect{p}_u\|^2_2 + \|\matr{Q}\|^2_F \right) \to \min_{\{ \matr{Q}, \vect{p}_u \}},
\end{equation}
where $\vect{a}_u$ is a binary indicator vector of length $N$ with value $a_{ui}$ in the $i$-th position, the matrix $\matr{C}_u\inR{N\times N}$ is diagonal, where the $i$-th element on the diagonal equals to $c_{ui}$, respectively.
Also, we use the following notations $\left\|\vect{x}\right\|^2_\matr{W}=\vect{x}^\top \matr{W} \vect{x}$ and $\| \matr{Q}\|_F^2$ is squared Frobenius norm of matrix $\matr{Q}$.

The standard matrix factorization model~\cite{Ammad_2019_FedCF} computes item relevance scores as $\mathbf{r}_{MF}(\matr{Q}, \vect{p}_u) = \matr{Q}\vect{p}_u$ that does not capture the sequential nature of the available data.
To capture the user's sequential patterns we introduce the following relevance score function $\vect{r}$ inspired by developments in~\cite{Jannach_2018_Eval_SBRS}:
\begin{equation}
\label{eq:prediction-rule}
    \vect{r}(\matr{Q}, \vect{p}_u) =
    \matr{Q}\vect{p}_u + \vect{h}_u(\matr{Q}) = 
    \matr{Q}\vect{p}_u + \diag{\matr{S}_u\matr {Q}\matr{Q}^\top},
\end{equation}
where the design matrix $\matr{S}_u$ encodes relative frequency weights corresponding to transition to app $i$ from some previous app $j$ in user $u$ history. 
The second term in~\eqref{eq:prediction-rule} depends only on the local transition statistics and therefore captures sequential behavior.

Elements of $\matr{S}_u$ are calculated based on user $u$ interaction history $\mathcal{H}_u^\ind{t}$ according to the following equation:
\begin{equation}
    \label{eq:design_matrix}
    \left(\matr{S}_u\right)_{ij}=
    \ddfrac{\sum_{k=2}^{|\mathcal{H}_u^\ind{t}|}\mathds{1}[j\rightarrow i]\mathds{1}[i_{k-1}^u=j] \mathds{1}[i_k^u=i]}{\sum_{k=2}^{|\mathcal{H}_u^\ind{t}|}\mathds{1}[i_{k-1}^u=i]}, 
    \mathds{1}[m=n] = \begin{cases} 1, & m = n\\ 0, & \text{otherwise}, \end{cases}
    \;
    \mathds{1}[j\rightarrow i] = 
    \begin{cases}
        1, & (j, i) \in \mathcal{H}_u^\ind{t}\\
        0,  & \text{otherwise},
    \end{cases}
\end{equation}
where $(j, i) \in \mathcal{H}_u^\ind{t}$ indicates that both $i$ and $j$ belong to the history $\mathcal{H}_u^\ind{t}$ and $i$ immediately follows $j$.
For example, let a user $u$ has the following history of interactions: $\mathcal{H}_u^\ind{t} = \left(a, b, c, a, a, b, a, c\right)$. 
Then, the resulting matrix $\matr{S}_u$ is of the size $3 \times 3$ and has the following form:
\begin{equation}
\matr{S}_u = \begin{blockarray}{cccc}
        a & b & c & \\
       \begin{block}{(ccc)c}
         \frac{1}{4} & \frac{2}{4}  & \frac{1}{4} & a \\
         \frac{1}{2} & 0 & \frac{1}{2} & b \\
         \frac{1}{1} & 0 & 0 & c\\
       \end{block}
     \end{blockarray}
\end{equation}
For instance, consider the $(a, b)$-th element of this matrix equal to $2/4$ since there are $2$ sequential co-occurrences of $(a, b)$ and 4 occurrences of $a$ in history.

The $i$-th element of $\vect{h}^{(t)}_u(\matr{Q})$ represents the average similarity between the target app $i$ and other apps weighted by the corresponding elements of the matrix $\matr{S}_u$ which encodes the sequential history~\cite{McAuley_2016_Fossil}:
\begin{equation}\label{eq:samf_group}
    \left(\vect{h}_u^\ind{t}\right)_i=
\sum_{j=1}^N\left(\matr{S}_u^\ind{t}\right)_{ij}\vect{q}_i^\top\vect{q}_j.
\end{equation}
Here we explicitly use the superscript $(t)$ to highlight that the duration of the considered period is important.
To improve capturing the global dynamics in app usage frequencies, one has to restrict a  period duration.
It can be done using a time window of the most recent events e.g. several weeks or months depending on the available data. 
Smaller~$t$ improves computational performance while larger $t$ may help discover longer-range correlations between events that do not immediately follow one another. 

Thus, we have introduced the novel relevance score function~\eqref{eq:prediction-rule} appropriate for sequential data and modified classical loss in the Matrix Factorization model.
We refer to the resulting objective function and relevance score function as Sequential Matrix Factorization model or \textbf{SeqMF}.

\subsubsection{Inference in SeqMF model}
The relevance score function~\eqref{eq:prediction-rule} is designed to facilitate proper learning of app co-occurrence statistics from a history of user actions.
However, prediction rule~\eqref{eq:prediction-rule} incurs high computational cost prohibitive in the inference step.
Therefore, in the inference step, we propose the following more lightweight relevance score function for user $u$ and the $i$-th app installed in the user device:
\begin{equation}
    \label{eq:sum_prediction}
\hat{r}(\vect{q}_i, \vect{p}_u; s_u) = \vect{q}_{i}^\top\vect{p}_{u} + \vect{q}_{i}^\top \vect{q}^{(s)}, \qquad  \vect{q}^{(s)} = \sum_{i_k \in s_u}\vect{q}_{i_k},
\end{equation}
where $s_u = (i_1, \dots, i_l), i_k \in A_u$ is a set of the latest $l$ apps user $u$ interacted with.
The first item $\vect{q}_{i}^\top\vect{p}_{u}$ is equal to the $i$-th element of the corresponding first item from~\eqref{eq:prediction-rule} and the second item $\vect{q}_{i}^\top \vect{q}^{(s)}$ is the approximation of the second item in~\eqref{eq:prediction-rule} that is responsible for capturing local patterns of user behavior.
Thus, we reduce the computational cost of the inference step while preserving the main property of the relevance score function simultaneously.




\subsubsection{Confidence-based weighting coefficients}
\label{sec:confidence}
The idea of the confidence-based weighting approach is to assign proper importance weights to discrepancies between predicted and ground-truth values of both observed and unobserved interactions during the training of a model. 
That way one can expect to learn better embeddings and capture underlying behavioral patterns more accurately. 
Popular weighting schemes~\cite{Koren_2008_ICF} assign small uniform weights on unobserved data 
and higher data-dependent weights on observed entries. 
Such schemes proved to be effective in practice for standard CF tasks~\cite{Koren_2008_ICF, Li_CF_Wights_2018}, where any available item can be recommended to any user.  
However, preliminary experiments showed that standard weighting schemes did not work in our setting.
The possible reason is the next app can be selected \emph{only} from the pool of installed apps rather than all possible apps.

Therefore, following the ideas on frequency-based weighting~\cite{He_FMF_2017}, we propose the following expression for confidence weights:
\begin{equation}
    c_{ui} = \frac{d^\gamma_{ui} + \alpha}{\sum_j d^\gamma_{uj} + \alpha N},
    \label{eq:rel_freq}
\end{equation}
where $d_{ui}$ is a launch frequency of item $i$ in the history of user $u$, and $\alpha \in [0; 1], \gamma \in [0; 1]$ are hyper-parameters. 
By modifying relative frequencies $d_{ui}$ with $\alpha$-additive (Laplace) smoothing~\cite{Valcarce_2016_Smoothing}
we make the confidence values on inactive apps a small non-zero constant, which is fully compatible with conventional weighted ALS-based approaches.
Note that nonzero $\alpha$ leads to an increase in computational cost due to implementation peculiarities.

\subsection{Federated learning of SeqMF model}
\label{sec::federated_framework}

Since the history of users' interactions is stored locally in users' devices it is natural to exploit the federated learning approach to solve~\eqref{eq:gen_loss}.
The federated learning approach specified for CF setup consists of two major intercommunicating blocks~\cite{ammad2019fedMF}: apps' embeddings $\matr{Q}$ located at the remote server and user embeddings $\vect{p}_u, \; u=1,\dots, M$ stored locally at the user devices (one vector per device).
The most recent version of matrix $\matr{Q}$ is distributed to the devices. Then, each device can update the corresponding user embedding using newly collected behavioral data and generate more accurate predictions. 
With some periodicity, some data from users' devices is sent back to the remote server to update the matrix $\matr{Q}$, and the process starts over.

The key ingredient of this scheme 
is the difference in the methods of updating users' and apps' embeddings.
As observed in~\cite{ammad2019fedMF}, the default ALS scheme to update matrix $\matr{Q}$ requires sending the entire user interactions history to the server, which poses a direct threat to data privacy.
To mitigate this problem, the authors use a hybrid optimization scheme that combines ALS updates for users' embeddings and gradient-based updates for apps' embeddings.

According to the ALS scheme, matrix $\matr{P}$ is updated in a row-wise manner as follows:
\begin{equation}
\label{eq:p_u}
\vect{p}_{u} = \left(\matr{Q}^\top \matr{C}_{u}\matr{Q} + \lambda \matr{I} \right)^{-1}\matr{Q}^\top  
\matr{C}_{u}
\left(\vect{a}_{u} - \vect{h}_u(\matr{Q})\right).
\end{equation}
We highlight that every user embedding $\vect{p}_u$ is updated \emph{privately on a user device} and \emph{in parallel} to support a distributed nature of federated learning.
Also, note that during such an update the influence of the long-term preferences encoded in $\vect{a}_u$ is compensated with the short-term sequential dynamics represented by $\vect{h}_u(\matr{Q})$. 
This is precisely the behavior we planned to achieve.

In contrast, the matrix $\matr{Q}$ is updated on the remote server with a gradient-based optimizer.
For example, classical gradient descent reads as follows: 
\begin{equation}
    \matr{Q} := \matr{Q} - \beta\,\ddfrac{\partial\mathcal{L}}{\partial \matr{Q}},
    \label{eq::q_update}
\end{equation}
where $\beta > 0$ is a learning rate, $\ddfrac{\partial \mathcal{L}}{\partial \matr{Q}}$ is full gradient of the introduced loss function~\eqref{eq:gen_loss} computed as follows:
\begin{equation}
    \label{eq:seqmf_Q_update}
    \ddfrac{\partial \mathcal{L}}{\partial \matr{Q}} =
    \sum_{u = 1}^{M} \matr{F}(u) + \lambda \matr{Q}, \qquad  
    \matr{F}(u) = 
    \matr{D}_u \vect{e}_N\vect{p}_u^\top + \left(\matr{D}_u\matr{S}_u + {\matr{S}_u}^\top\matr{D}_u\right)\matr{Q},
\end{equation}
where $\matr{D}_u = \diag{\matr{C}_{u}\left(\vect{r}(\matr{Q}, \vect{p}_u) - \vect{a}_{u}\right)}$, $\vect{e}_N$ is a vector of ones of size $N$ and $\mathrm{diag}(w)$ is a diagonal matrix with vector $w$ in the diagonal.

To compute $\ddfrac{\partial \mathcal{L}}{\partial \matr{Q}}$ on a server side we need only values $\matr{F}(u)$ from~\eqref{eq:seqmf_Q_update}. 
These values can be gathered from clients on the remote server.
However, in practice, not all the clients may send $\matr{F}(u)$ by request.
This issue is treated by using a stochastic gradient estimator:
\begin{equation}
    \ddfrac{\partial \mathcal{L}}{\partial \matr{Q}} \approx \sum\limits_{u \in U_b}\matr{F}(u) + \lambda \matr{Q},
    \label{eq::stoch_grad}
\end{equation}
where $U_b$ is a set of users that can transmit $\matr{F}(u)$.
After that, a stochastic optimizer like SGD with momentum~\cite{Qian_1999_Momentum} or Adam~\cite{Kingma2014} can be used to update matrix $\matr{Q}$ in the remote server.

Although the described hybrid scheme of federated learning requires transmitting $\matr{F}(u)$ from users' devices to the remote server, this data does not contain explicit user interaction history, so it is less sensitive.
Nevertheless, it is still susceptible to certain types of attacks, so we make this transmission private. We aggregate the anonymized data in the remote server and use the perturbed gradient estimate $\bar{\matr{F}}$ to update apps' embeddings $\matr{Q}$.  
To ensure stronger privacy of the transmitted data, we propose a new privacy-preserving mechanism described  in the next section.

\section{Privacy mechanism in the SeqMF model}

\subsection{Local Differential Privacy}
\label{subsec:LDP}
\begin{figure}[!h]
    \centering
    \includegraphics[scale=0.4]{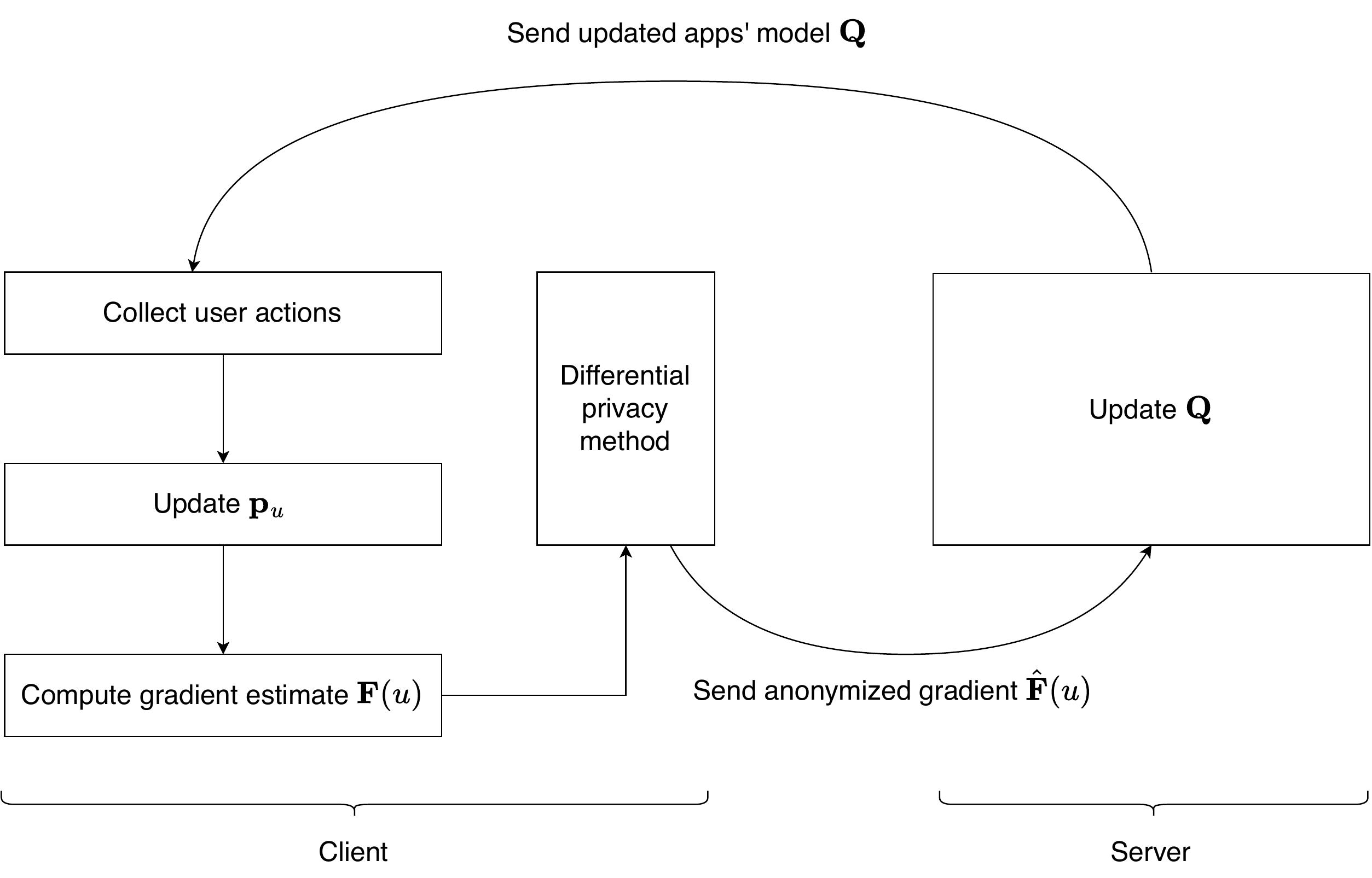}
    \caption{The scheme of communications between server and clients and what data transmission is performed.}
    \label{fig:one_client}
\end{figure}

Incorporation of privacy mechanism into the model based on matrix factorization is straightforward and natural. 
Figure~\ref{fig:one_client} illustrates  the scheme of private communications between one client and the remote server. 
In this scheme, user embeddings $\vect{p}_u$ and the history of users' actions $\mathcal{H}^{(t)}_u$ are stored only on the device to preserve the privacy of sensitive information.
However, to update the apps' embeddings in matrix $\matr{Q}$ users compute gradients $\matr{F}(u)$ with private data.
Transmission of these gradients to the remote server can lead to the leakage of private data. 
Thus, to maintain privacy throughout the learning process the additional privacy-preserving mechanism is used before the transmission of the gradients. 

In general, there are several approaches to protect sensitive data: secure Multi-Party Computation~(MPC)~\cite{Yao80MPC}, Homomorphic Encryption~(HE)~\cite{HE_Rey}, Trusted Execution Environment~(TEE)~\cite{intel_cpu,arm_cpu,sact_cpu}, Differential Privacy~(DP)~\cite{DP_centr,DP_centr_book,DP_centr_2}. 
However, TEE imposes a large number of restrictions on the hardware level. 
MPC and HE significantly increase the computational complexity or require the trusted third party~\cite{fl_main}.
So, further, we consider only the local differential privacy~($\epsilon$-LDP) technique, which gives strong privacy guarantees without large additional overhead on computational complexity and communication cost~\cite{fl_main}. 
Also, the $\epsilon$-LDP mechanism does not require a trusted third party, unlike central differential privacy algorithms~(CDP)~\cite{DP_centr_book}. 

\begin{definition}
A randomized mechanism $\mathcal{A}$ is $\epsilon$-locally differentially private ($\epsilon$-LDP) if and only if for any two inputs $v,\ v^{\prime}$ and any output $y \in range(\mathcal{A})$ the following holds:
\begin{equation}
    \mathbb{P}(\mathcal{A}(v) = y) \leq e^{\epsilon}\mathbb{P}(\mathcal{A}(v^{\prime}) = y),
\end{equation}
where $\epsilon$ is called \emph{the privacy budget}. 
The smaller the privacy budget is, the better protection of sensitive data is. 
\end{definition}

The privacy-preserving matrix factorization methods have already been considered in studies.
In particular, to perturb the gradients before transmission, the $k$-Harmony mechanism was incorporated in gradient aggregation privacy-preserving algorithm~\cite{Minto_2021_FedCF,shinface}. 
Below we discuss the drawbacks of this mechanism and propose its modification for our setting called \emph{QHarmony mechanism}. 
In addition, we discuss Laplace mechanism~\cite{DP_centr_book}, Kashin mechanism~\cite{Chen2020BreakingTC}, and their drawbacks in the next section.

\subsection{QHarmony mechanism}
\label{sec::q_harmony}
Consider gradient matrix $\matr{F}(u) = [f_{ij}(u)] \in \mathbb{R}^{N\times d}$ from user $u$, where $N$ is a number of apps, $d$ is a dimension of the embeddings.
Further, we assume that $f_{i j}(u) \in [-1, 1],\ i \in \overline{1,N}, \ j \in \overline{1,d}, \ u \in \overline{1,M}$.  
Since we can normalize the computed gradients, this assumption easily holds. 
The perturbed gradient matrix $\hat{\matr{F}}(u)$ is computed by the $\epsilon$-LDP mechanism $\mathcal{A}$ from $\matr{F}(u)$ on the device.
Then, a user transmits $\hat{\matr{F}}(u)$ to the remote server with privacy guarantees. 
On the server side, the received perturbed gradients $\hat{\matr{F}}(u), \; u \in U_b$ are aggregated and, if needed, the additional transformation is performed on the aggregated update. 
According to the standard aggregation scheme for mean estimation, $\bmatr{F} = [\bar{f}_{i j}]$ is computed as follows: 
\begin{equation}
    \bmatr{F} = \frac{1}{|U_b|}\sum\limits_{u \in U_b} \hat{\matr{F}}(u). 
    \label{eq::grad_priv_average}
\end{equation}
This way of aggregating gradients aims to reduce the noise from the received gradients $\hat{\matr{F}}(u)$ and therefore a sufficiently large number of users $M$ is needed. 
However, in our setup the number of users is approximately the same as the number of elements in the gradient matrix, thus, the baseline $\epsilon$-LDP mechanisms are not appropriate for such a setting. 
In particular, 
\begin{itemize}
    \item $k$-Harmony mechanism makes perturbed gradients sparse, since only $k$ elements are transmitted from every user.
    The choice of $k$ is based on the requirements of low privacy budget $\epsilon$ and high privacy budget per transmitted element $\frac{\epsilon}{k}$.
    The higher the latter budget is, the less noisy perturbed gradient element $\hat{f}_{ij}$ is.
    Therefore, $k$ should be small, e.g. $k < 10$.
    To compute $\bar{f}_{ij}$ only approximately $\frac{k M}{N d} \approx k$ values are summed and since $k$ is small the remote server can not estimate $\bar{f}_{ij}$ sufficiently accurately. 
    \item According to the Laplace mechanism~\cite{DP_centr_book}, the variance of the gradient perturbation is equal to $\frac{4 N^2 d^2 }{\epsilon^2}$, which is a large number for low privacy budget, e.g.  $\epsilon < 3$. 
    At the same time, the standard deviation of aggregated gradient element $\bar{f}_{ij}$ is $\frac{2 \sqrt{2} N d}{\epsilon M} \approx \frac{2 \sqrt{2}}{\epsilon} $.
    In the case of a low privacy budget, this standard deviation is comparable with the element $\bar{f}_{ij}$. 
    Therefore, the estimate $\bmatr{F}$ is inaccurate with the Laplace mechanism, too.
    \item Although, the Kashin mechanism has the lowest asymptotic variance upper bound for the aggregated gradient~$\bmatr{F}$ compared with other mechanisms, it has excessive computational costs. 
    Also, if the hidden constant in the asymptotic estimate is large, the Kashin mechanism gives an inaccurate estimate $\bmatr{F}$.
\end{itemize} 

Thus, in the considered setting, the baseline privacy-preserving mechanisms estimate the average gradient~\eqref{eq::grad_priv_average} with a large variance, that leads to poor performance.
To improve overall performance and preserve moderate computational cost, we modify $k$-Harmony mechanism as follows.
Since this mechanism can be considered as a quantization of $\bar{\matr{F}}$, we take the aggregation approach from the quantization methods.
Thus, our privacy mechanism consists of the following steps. 
Firstly, user samples uniformly $k$ different pairs of indexes $(i_l,j_l)$, where $i_l \in \overline{1,N}$, $j_l \in \overline{1,d}$ and $ l \in \overline{1,k}$.
Then, user assigns zeros to all elements of $\hat{\matr{F}}(u)$ except $\hat{f}_{i_l j_l}(u),\ l \in \overline{1,k}$, which are computed in the following way: 
\begin{equation}
\mathbb{P}[\hat{f}_{i_l j_l}(u) = x] = \begin{cases} 
\frac{f_{i_l j_l}(u) (e^{\epsilon/k} - 1) + e^{\epsilon/k} + 1}{2(e^{\epsilon/k} + 1)} , \ \text{if} \ x = 1 \\ 
- \frac{f_{i_l j_l}(u) (e^{\epsilon/k} - 1) + e^{\epsilon/k} + 1}{2(e^{\epsilon/k} + 1)} , \ \text{if} \ x = -1.
\end{cases}
\end{equation} 
In the final step, user transmits $k$ triples $(\hat{f}_{i_l j_l}(u), i_l, j_l)$ and $f_{\max}(u) = \max\limits_{i,j} f_{i j}(u)$ to the remote server.
Further, we observe that the aggregated gradient $\bmatr{F}$ on the server is a quantized representation of the non-private gradient estimate~\eqref{eq::stoch_grad}. 
Based on this observation, we replace the baseline aggregated formula~\eqref{eq::grad_priv_average} with the following expression: 
\begin{equation}
    \bar{\matr{F}} = \max_u f_{\max}(u) \sum\limits_{u \in U_b} \frac{\hat{\matr{F}}(u)}{\max\limits_{i j} \sum\limits_{u \in U_b} \left[\hat{f}_{i j}(u) > 0\right]},
\end{equation}
where $f_{\max}(u) = \max\limits_{i j} f_{ij}(u)$ and $f_{\max}(u)$ is transmitted to server by every user. 
We called this new privacy-preserving algorithm \textit{QHarmony} and summarize it for the user and server sides in Algorithms~\ref{alg:qharmonyclient} and~\ref{alg:qharmonyserver}, respectively.

\noindent
\begin{minipage}[t]{.48\textwidth}
\begin{algorithm}[H]
\caption{QHarmony mechanism: Client side}
\label{alg:qharmonyclient}
\begin{algorithmic}[1]
\Require{$\matr{F}(u)$ is the computed gradient on the user $u$ device, $\epsilon$ is a privacy budget, $k$ is a number of transmitted non-zero values.}
\Ensure{$\mathcal{I}(u)$ is set of $k$ triples $(\hat{f}_{i_l j_l}(u), i_l, j_l)$ corresponding to each non-zero position in the perturbed gradient $\hat{\matr{F}}(u)$ on the $u$-th user device, $f_{\max}(u)$ is the maximum element of $\matr{F}(u)$.}
    \Function{QHarmonyC}{$\matr{F}(u)$, $\epsilon$, $k$}
        \State Compute $f_{\max}(u) = \max\limits_{i j} f_{i j}(u)$
        \State Initialize: $\mathcal{I}(u) = \emptyset$
        \State Initialize: $\mathcal{P} = \{i\}_{i=1}^N \times \{j\}_{j=1}^d $
        \For{$l = 1,\ldots,k$}
            \State Sample uniformly at random $(i_l,j_l)$ from $\mathcal{P}$
            \State Sample random number $b \sim Bernoulli\left(\frac{f_{i_l j_l}(u) (e^{\epsilon/k} - 1) + e^{\epsilon/k} + 1}{2(e^{\epsilon/k} + 1)}\right)$
            \State $\hat{f}_{j_l i_l}(u) = 2b - 1$
            \State $\mathcal{P}$ = $\mathcal{P}(u)$ / $\{(i_l, j_l)\}$
            \State $\mathcal{I}$ = $\mathcal{I}(u) \cup \{(\hat{f}_{i_l j_l}(u), i_l, j_l)\}$
        \EndFor
        \\
        \Return $\mathcal{I}(u)$, $f_{\max}(u)$
    \EndFunction
\end{algorithmic}
\end{algorithm}
\end{minipage}
\hfill
\begin{minipage}[t]{.48\textwidth}
\begin{algorithm}[H]
\caption{QHarmony mechanism: Server side}
\label{alg:qharmonyserver}
\begin{algorithmic}[1]
\Require{For every user $u \in U_b$, $\mathcal{I}(u)$ is a set of $k$ triples $(\hat{f}_{i_l j_l}(u), i_l, j_l)$ corresponding to each non-zero element in the perturbed gradient $\hat{\matr{F}}(u)$ from the $u$-th user, $f_{\max}(u)$ is the maximum element of $\matr{F}(u)$.}
\Ensure{$\bmatr{F}$ is the resulting perturbed gradient in the remote server for the matrix $\matr{Q}$ update.}
    \Function{QHarmonyS}{$\mathcal{I}(u)$,$f_{\max}(u)$, $u \in U_b$}
        \State Initialize: $\matr{S} = \{0\}^{N\times d}$, $\matr{Z} = \{0\}^{N\times d}$

        \For{$u \in U_b$}
            \For{$(\hat{f}_{i_l j_l}(u), i_l, j_l) \in \mathcal{I}(u)$}
                \State $s_{i_l j_l} = s_{i_l j_l} + \hat{f}_{i_l j_l}(u)$ \Comment{Aggregate gradients}
                \State $z_{i_l j_l} = z_{i_l j_l} + [\hat{f}_{i_l j_l}(u) > 0]$ 
                \EndFor
            \EndFor
        \State $z_{\max} = \max\limits_{i j} z_{i j}$
        \State $\bmatr{F} =  \frac{\max\limits_u f_{\max}(u)}{z_{\max}}\matr{S}$
        \\
    \Return $\bmatr{F}$
    \EndFunction
\end{algorithmic}
\end{algorithm}
\end{minipage}

\section{Experiments}
\label{sec:experiments}
\subsection{Data}
\label{sec:data}
Experiments were carried out on two mobile app usage datasets that are publicly available. 
LSApp dataset~\cite{AliannejadiTOIS21} is a collection of cross-app mobile search queries where the authors collected sequential app usage events of the users.
It contains $N=87$ unique apps from $M=292$ users. 
In App Usage dataset~\cite{Yu_App_Usage_2018}, each entry contains an anonymized user identification, timestamps of HTTP request or response, etc. It contains $N=2000$ unique apps from $M=1000$ users. 
We noticed an excessive redundancy in both considered datasets: the same app's launch event can occur multiple times in less than 3 seconds. We believe, this is non-representative of real user behavior and collapse such cases into a single event. 

Moreover, we split user actions into sessions based on a time interval between consequent app launches. 
If in the current session the time delay after the last seen event exceeds a predefined threshold (15 minutes), a new session starts. The average number of sessions by users for LSApp and for App Usage is 109 and 58, respectively. 
Session splitting is only used during evaluation and does not affect the training process. During training, we consider only the sequential nature of data and represent all user history as a long session. 

\subsection{Evaluation methodology}
\label{sec:evaluation}
\subsubsection{Top-$n$ prediction}
The main goal is to predict the next app that is available on a user's device, not in the entire app store. 
Hence, we restrict the possible predictions to the apps installed on the user's device. 
For example, if a user history contains interactions with 5 unique apps, we will require a model to generate prediction scores for these 5 apps only. 
Then, for top-$n$ recommendations with $n=3$, we select 3 apps with the highest predicted scores among those 5 apps. 
Note that the number of apps $|A_u|$ in user $u$ history can be lower than $n$. 
Hence, we compute top-$\min(|A_u|, n)$ predictions and still assign the result to top-$n$ metrics.

\subsubsection{Evaluation protocol}
Within each session we utilize \emph{iterative revealing scheme}~\cite{Ludewig_2018_EvalSess} for evaluating the quality of predictions.
This scheme means that for a particular user $u$ and session $s$ we go iteratively through apps in $s$ and predict the next app taking all the previous apps into account.
Hence, for a session of length $l$ we successively generate $l-1$ predictions. 
We use both relevance and ranking-related metrics like hit-rate (HR$@n$)~\cite{Zhang_NIRSA_2018}, mean reciprocal rank (MRR$@n$)~\cite{Nikolakopoulos_eigenrec_2015}, and normalized discounted cumulative gain (NDCG$@n$)~\cite{Nikolakopoulos_eigenrec_2015}, where $n$ corresponds to the number of recommended items.
Since our setting slightly differs from the standard one, we modify these metrics respectively but preserve their properties.
In particular, the averaging is performed not only by the number of users but also by sessions.

\subsection{Methods for comparison}

\textit{Matrix Factorization.} This is the original non-sequential MF-based approach proposed in~\cite{ammad2019fedMF}.

\textit{Sequential Rules.}
We use a Sequential Rules (SR) predictor~\cite{Ludewig_2018_EvalSess} based on the co-occurrence frequency of apps going immediately one after another in user history. 
This is a direct modification of the standard item-to-item approach based on the Markov Chain assumption.
After the statistics of such pair-wise sequential co-occurrences are collected, we use them directly without normalization to assign prediction scores for the next item. 
Note that there are two possible variations of the SR model depending on the way data is collected. 
The model can be learned either in a collaborative or in an isolated (on-device) manner. 
We denote the former as \emph{SR} and the latter as \emph{SR-od}.

\textit{Most Recently Used.} MRU predictor assigns scores based on the recency of items in the current session. 
To comply with evaluation protocol (see \sref{sec:evaluation}), items that do not belong to requested items get zero scores. 
The non-zero scores are lower for older items and higher for newer ones, which recent session items to the top in generated predictions.

\textit{Most Frequently Used.} MFU predictor computes item popularity scores based on events frequencies for each user individually. 
These scores are used to select  the top most frequently used items for prediction.

\textit{Random.} Finally, this predictor assigns random scores to all requested items.

It is worthwhile to say that despite there are more complex models like~\cite{Gang2019AppUsage2Vec,Wan2021DeepAPP} \emph{the main purpose of this study is to develop a privacy-preserving method for the on-device next app prediction that outperforms well-known model in real-world scenario}. 
Therefore, approaches based on the privacy-preserving federated learning of neural networks are out of the scope of this study and the subject of future work.

\subsection{Experimental methodology}
\label{sec:exp_methodology}

\subsubsection{Static environment}
\label{sec:static_env}
The static environment~\cite{McAuley_2016_Fossil} aims to evaluate how well the SeqMF model is able to process sequential data compared to standard baselines.
In particular, in this environment, the standard splitting on the train, test, and validation subsets is used for training and evaluation of the app prediction model.
The validation subset is used to tune hyperparameters.
In particular, we tune a dimension of users' and apps' embeddings $d$, the learning rate of optimizer $\beta > 0$ from~\eqref{eq::q_update}, regularization coefficient $\lambda > 0$, and smoothing parameters $\alpha$ and $\gamma$ from~\eqref{eq:rel_freq}.
In the LSApp dataset, the train set is the first 222 days, the  validation set is the next 7 days, and the test subset is the last 14 days.
For App Usage data, we assign the first 6 days to the training subset, the next day is assigned to the validation subset, and the last day is assigned to  the test subset.

\subsubsection{Dynamic environment}
\label{sec:dynamic_env}
Dynamic environment corresponds to the behavior of the model in a real-world setting, where the model is built into the mobile OS and starts working on a new device. 
To achieve high performance, the model has to adapt to user behavior quickly and starts generating reasonable predictions.
Dynamic environment assumes that we have pre-trained apps' embeddings $\matr{Q}$ and randomly initialized users' embeddings $\vect{p}_u$ stored on devices.
Then embeddings are updated from the online data stream.
For example, consider a data stream that is generated over 5 days. 
In a dynamic environment, we evaluate the initialization of the SeqMF model with the data from the first day.
Then, we update embeddings from this data and evaluate the model on the data from the second day, and then use the data from the second day to update embeddings.
This procedure continues till the fifth day. 
In the end, we get five numbers that show the dynamic behavior of the model's quality in terms of the chosen metrics.

\begin{figure}[h!]
\begin{subfigure}{.5\textwidth}
  \centering
  \includegraphics[width=.9\linewidth]{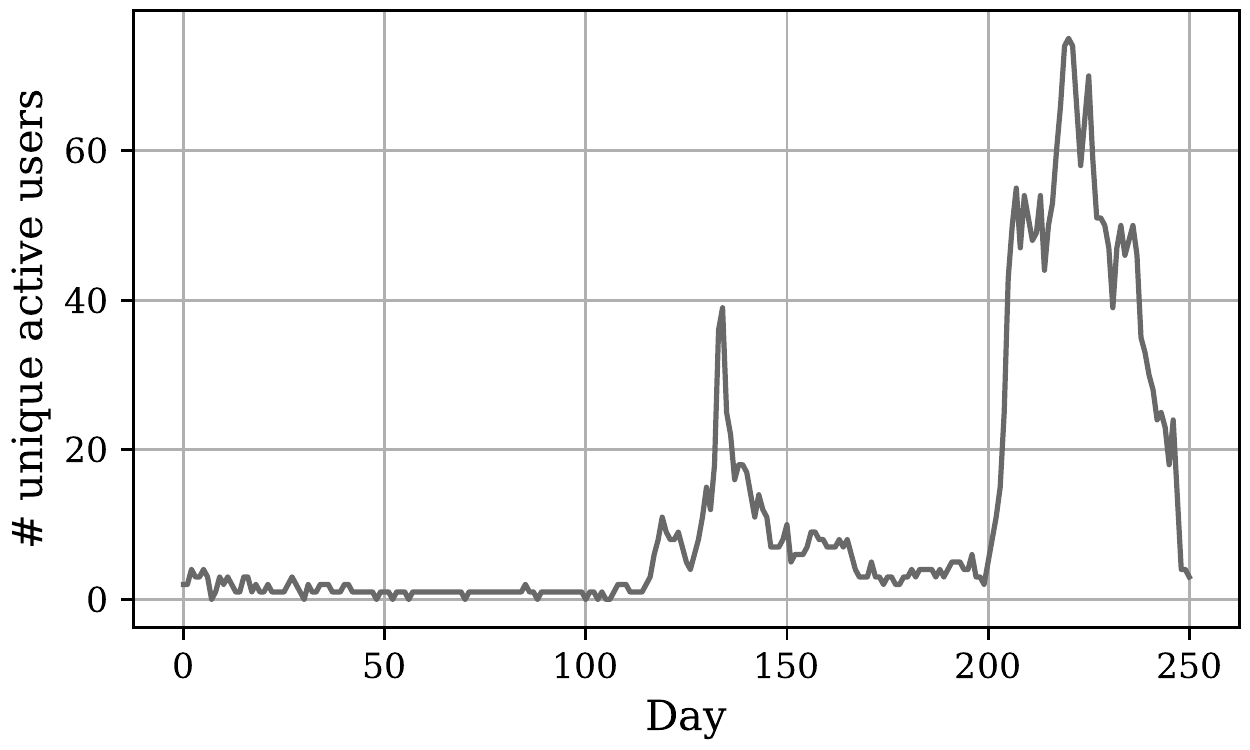}
\end{subfigure}%
\begin{subfigure}{.5\textwidth}
  \centering
  \includegraphics[width=.9\linewidth]{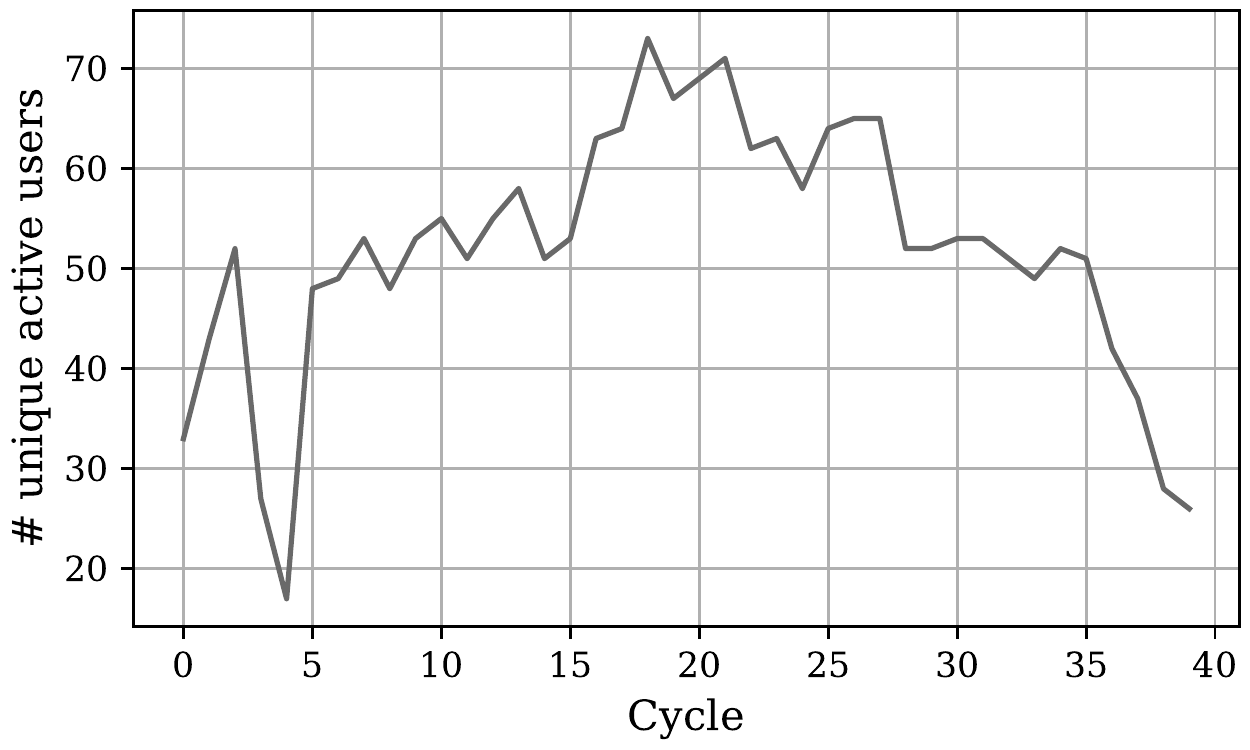}
\end{subfigure}
    \caption{
        Comparison of the original statistics from the LSApp dataset (left) and statistics after rearrangement into cycles (right).}
    \label{fig:dynamic_preprocessing}
\end{figure}

One more important factor in the dynamic environment setting is the distribution of users' activities over time.
If this distribution is unbalanced, then the learning process is unstable due to a noisy gradient estimated on small fractions of data.
We observe the exact such case in the LSApp dataset, see~\fref{fig:dynamic_preprocessing}~(left).
To make the considered dataset balanced, we artificially rearrange users' actions in the \emph{cycles} such that the number of active users is almost the same in every cycle, see~\fref{fig:dynamic_preprocessing}~(right).
Each cycle may contain one or several days, depending on user activity.
At the same time, the order of users' actions was preserved during such rearrangement.
In order to get pre-trained apps' embeddings $\matr{Q}$ we take the first cycle for training the SeqMF model. 
The further cycles are composed of a data stream for the dynamic environment.


\subsection{Results}
\label{subsec:results}
\subsubsection{Static environment}
\label{subsec:static_results}
A comparison of the proposed SeqMF model and competitors described above in the static environment is presented in Table~\ref{tab:lsapp} for LSApp and in Table~\ref{tab:app_usage} for App Usage data.
We see that SeqMF significantly outperforms its non-sequential predecessor MF on both datasets, which is expected given the sequential nature of the data.
However, it underperforms the simple SR-od model on LSApp dataset and SR model on App Usage data in all the metrics.
The SR models are known to be generally strong baselines and perform well on various datasets~\cite{Ludewig_2018_EvalSess}. 
It should be noted, that there are at least two reasons for such results. 
Firstly, there are a lot of repetitive events in the datasets, including long sequences consisting of the same app. 
Thus, even despite our attempts to filter out such anomalies, they induce a strong bias towards duplicated events. 
Another reason is that once enough statistical data is collected, the sequential patterns become close to stationary and easier to uncover.
Thus, we present comparison results in a dynamical environment that corresponds to the real-world scenarios in~\sref{subsec:dynamic_results}.

\begin{table}[!h]
\caption{Static evaluation results on the LSApp dataset.}
\label{tab:lsapp}
\centering
\begin{tabular}{lrrrrrrrrr}
\toprule
Models & HR@1 &  HR@3 &  HR@5 &  MRR@1 &  MRR@3 &  MRR@5 &  NDCG@1 &  NDCG@3 &  NDCG@5 \\
\midrule
Random &               0.089 & 0.239 & 0.365 &  0.089 &  0.153 &  0.181 &   0.089 &   0.175 &   0.227 \\
MRU &                  \textbf{0.618} & \underline{0.823} & 0.849 &  \textbf{0.618} &  \textbf{0.716} &  \textbf{0.722} &   \textbf{0.618} &  \textbf{0.744} &   \underline{0.754} \\
MFU &                  0.403 & 0.669 & 0.777 &  0.403 &  0.516 &  0.541 &   0.403 &   0.555 &   0.600 \\
SR &                   0.610 & 0.801 & 0.863 &  0.610 &  0.695 &  \underline{0.709} &   0.610 &   0.723 &   0.748 \\
SR-od &               \underline{0.612} & \textbf{0.824} & \textbf{0.893} &  \underline{0.612} &  \underline{0.707} &  \textbf{0.722} &   \underline{0.612} &   \underline{0.737} &   \textbf{0.765} \\
MF &                   0.379 & 0.649 & 0.784 &  0.379 &  0.497 &  0.527 &   0.379 &   0.536 &   0.591 \\
\midrule
SeqMF &    0.522 & 0.797 & \underline{0.871} &  0.522 &  0.643 &  0.660 &   0.522 &   0.683 &   0.713 \\
\bottomrule
\end{tabular}
\end{table}
\begin{table}[!h]
\caption{Static evaluation results on the App Usage dataset.}
\label{tab:app_usage}
\centering
\begin{tabular}{lrrrrrrrrr}
\toprule
Models &               HR@1 &  HR@3 &  HR@5 &  MRR@1 &  MRR@3 &  MRR@5 &  NDCG@1 &  NDCG@3 &  NDCG@5 \\
\midrule
Random &              0.029 & 0.093 & 0.152 &  0.029 &  0.056 &  0.069 &   0.029 &   0.065 &   0.090 \\
MRU &                 0.393 & \underline{0.660} & \underline{0.707} &  0.393 &  \underline{0.516} &  \underline{0.527} &   0.393 &   \underline{0.553} &   \underline{0.572} \\
MFU &                 0.229 & 0.436 & 0.546 &  0.229 &  0.319 &  0.344 &   0.229 &   0.349 &   0.394 \\
SR &                  \textbf{0.477} & \textbf{0.664} & \textbf{0.746} &  \textbf{0.477} &  \textbf{0.559} &  \textbf{0.578} &   \textbf{0.477} &   \textbf{0.586} &   \textbf{0.620} \\
SR-od &              \underline{0.414} & 0.591 & 0.659 &  \underline{0.414} &  0.493 &  0.508 &   \underline{0.414} &   0.518 &   0.546 \\
MF &                  0.227 & 0.426 & 0.520 &  0.227 &  0.314 &  0.336 &   0.227 &   0.343 &   0.382 \\
\midrule
SeqMF &               0.402 & 0.615 & 0.686 &  0.402 &  0.497 &  0.513 &   0.402 &   0.527 &   0.557 \\
\bottomrule
\end{tabular}
\end{table}

\subsubsection{Dynamic environment: regular update of the user embeddings is important}\label{subsec:dynamic_results}
In this section, we investigate the impact of long-term and short-term components in prediction rule~\eqref{eq:prediction-rule} on model performance. We carry out experiments using setup from~\sref{sec:dynamic_env}.

We consider three regimes of the SeqMF model training: Full, Rare, and Global.
In these regimes, apps' embeddings~$\matr{Q}$ are updated every 10 cycles for the LSApp dataset and every two cycles for the App Usage dataset. 
In the Full regime, users' embeddings $\vect{p}_u$ are updated every cycle. 
In the Rare regime, users' embeddings~$\vect{p}_u$ are updated only with the updates of apps' embeddings~$\matr{Q}$. 
In the Global regime, users' embeddings $\vect{p}_u$ are never updated and remain the same after random initialization.
SR-od model is re-trained in each cycle from the data collected in previous cycles.

To evaluate the performance of the model in a dynamic environment, we introduce an additional metric which is denoted by $\delta \mathrm{HR}@5$.
This metric is equal to the difference between $\mathrm{HR}@5$ computed from the target training regime and $\mathrm{HR}@5$ computed from the SeqMF model trained in the Full regime since we assume it is the most representative. 
In the plots below, we provide the cumulative $\delta \mathrm{HR}@5$ over the data stream to illustrate the impact of the users' embeddings updates schedule on the SeqMF model performance.
We also plot the performance of the SR-od model since it shows high performance in the static environment (see Table~\ref{tab:lsapp} and~\ref{tab:app_usage}) and uses only the local data from the device.
The latter feature makes the SR-od model appropriate in the considered setting which is close to the real-world scenario.

\begin{figure}[h!]
\begin{subfigure}{.5\textwidth}
  \centering
  \includegraphics[width=.9\linewidth]{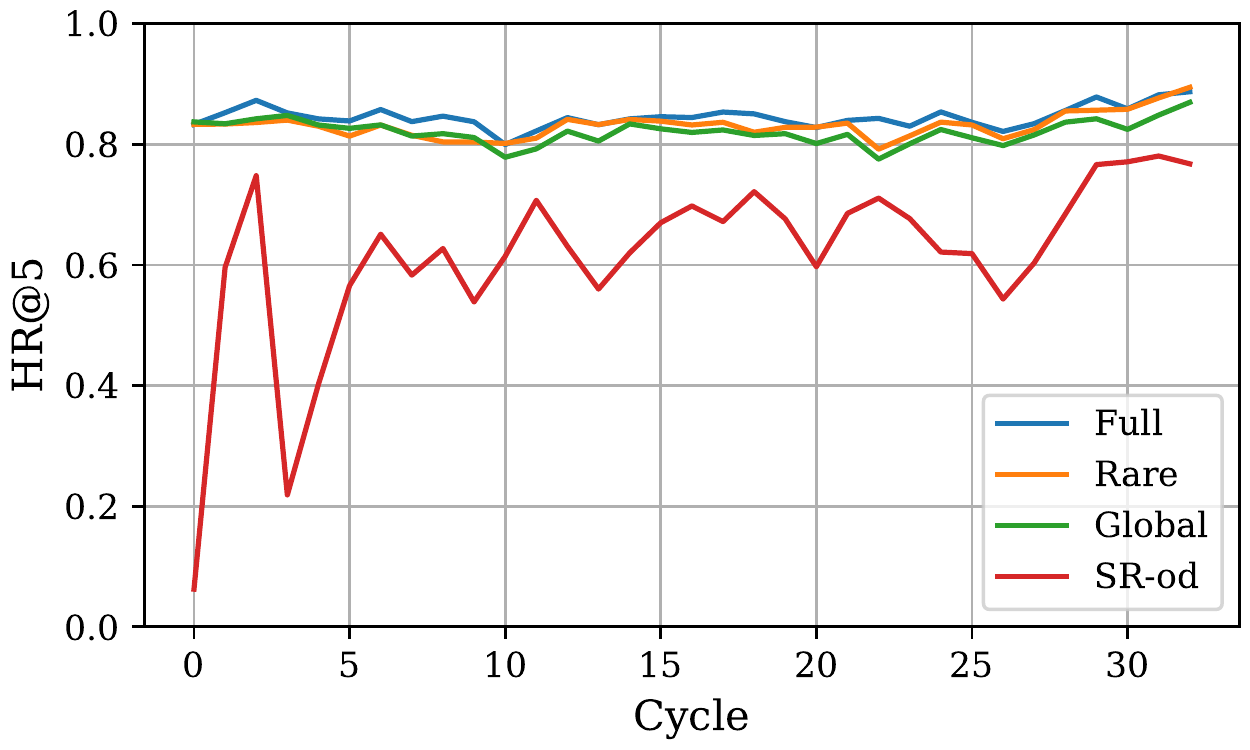}
\end{subfigure}%
\begin{subfigure}{.5\textwidth}
  \centering
  \includegraphics[width=.9\linewidth]{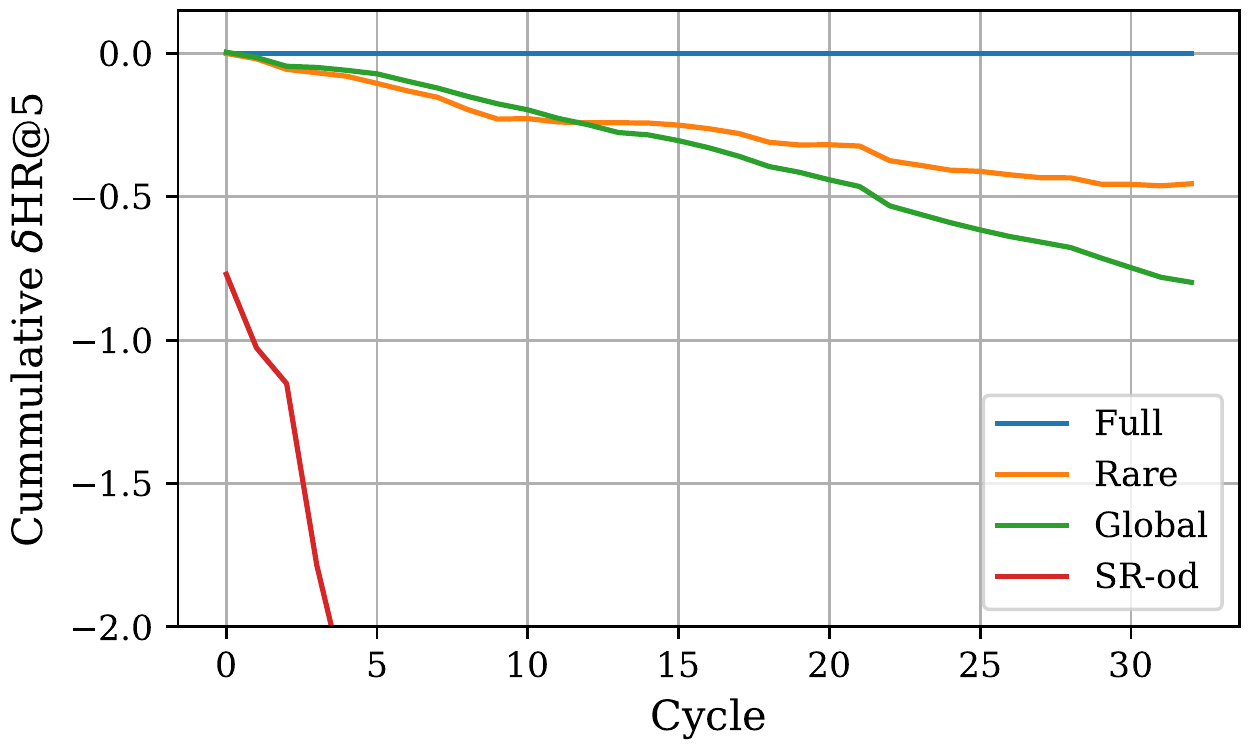}
\end{subfigure}
    \caption{
        \emph{LSApp dataset.}
$\mathrm{HR}@5$ metric (left) and cumulative $\delta\mathrm{HR}@5$ (right) for the considered training regimes of the SeqMF model and SR(od) model in the dynamical environment.
The SeqMF model in the considered training regimes outperforms SR-od model in dynamical environment.
The right plot shows that regular updates of user embeddings improve performance.}
    \label{fig:ablation_study_lsapp}
\end{figure}

Figure~\ref{fig:ablation_study_lsapp} illustrates that the SeqMF model \emph{outperforms the SR-od model on LSApp data in the dynamic environment}.
This observation confirms our hypothesis about the reasons for the high performance of SR-based models in the static environment.
As we can see, if we do not update users' embeddings on the device at all (Global regime), then the overall performance of the SeqMF model decreases. 
However, if we add just rare updates of users' embeddings (Rare regime), we can see that the performance increases compared to the SeqMF model in the Global regime. 
Thus, we demonstrated experimentally that updates of users' embeddings $\vect{p}_{u}$ are essential to preserving the high performance of the SeqMF model. 

\begin{figure}[h!]
\begin{subfigure}{.5\textwidth}
  \centering
  \includegraphics[width=.9\linewidth]{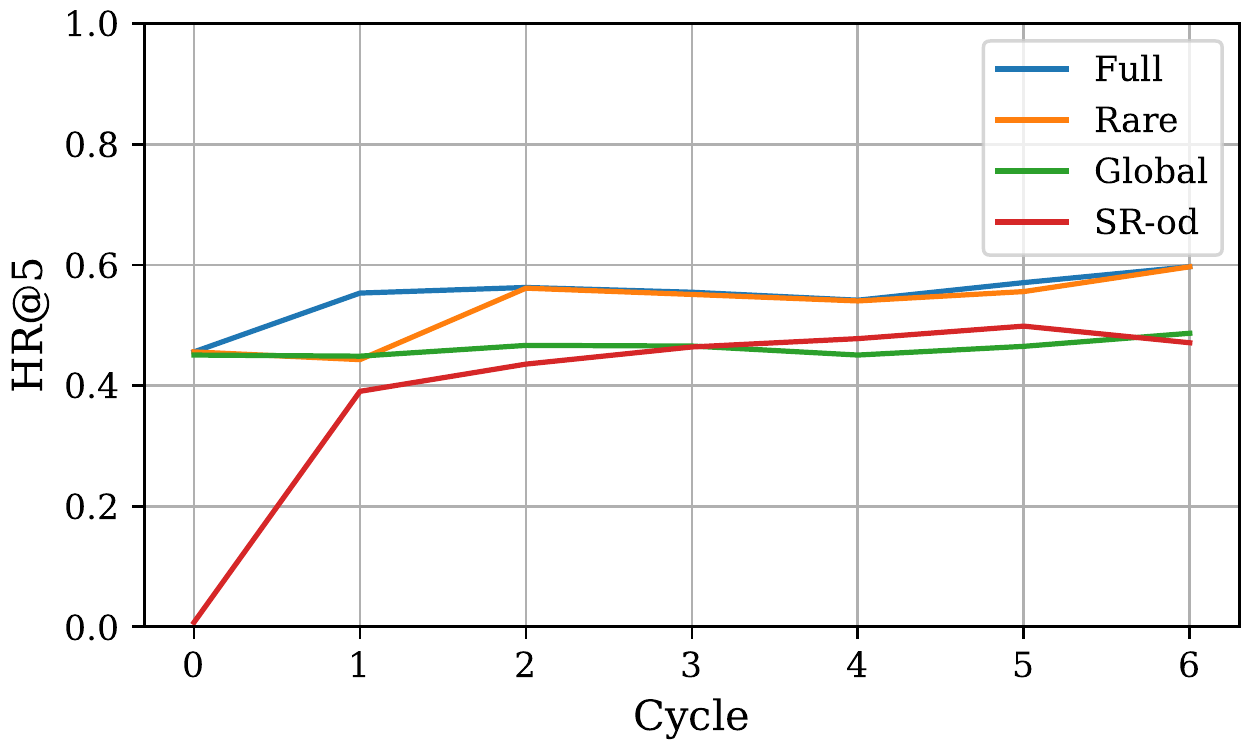}
\end{subfigure}%
\begin{subfigure}{.5\textwidth}
  \centering
  \includegraphics[width=.9\linewidth]{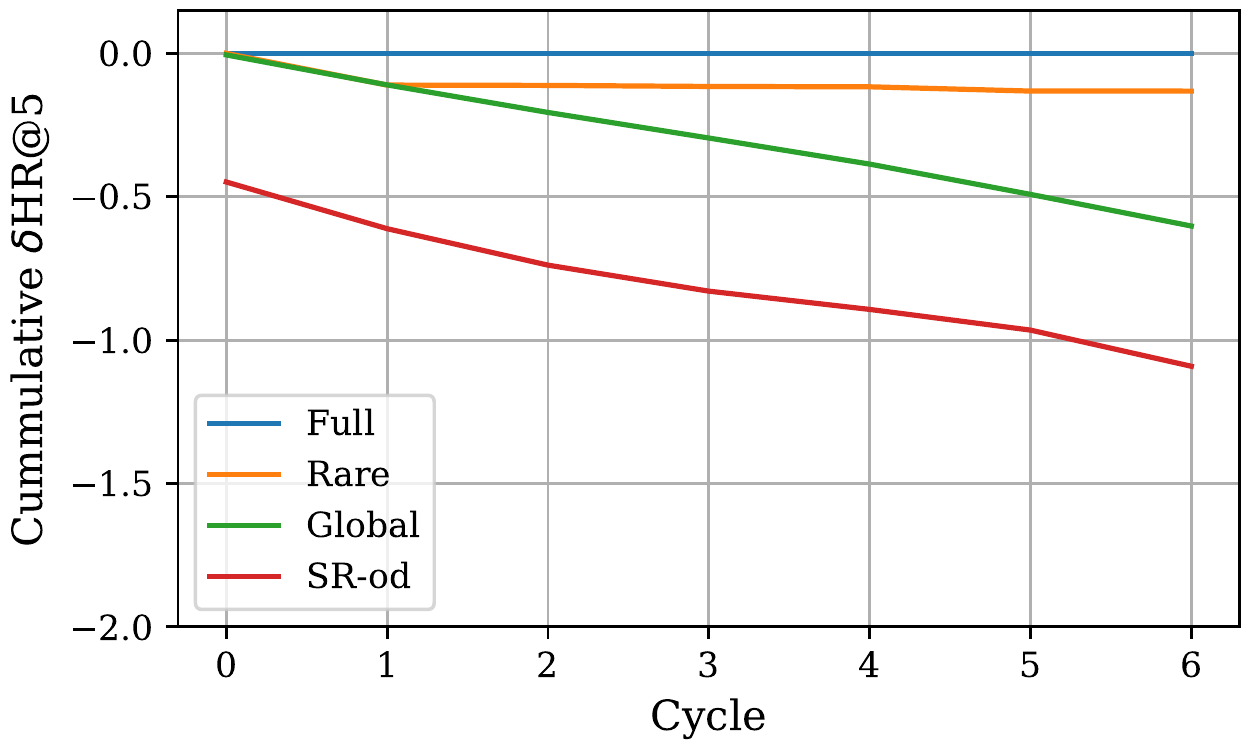}
\end{subfigure}
    \caption{
        \emph{App Usage dataset.}
        $\mathrm{HR}@5$ metric (left) and cumulative $\delta\mathrm{HR}@5$ (right) for the considered training regimes of the SeqMF model and SR(od) model in the dynamical environment.
The SeqMF model in the considered training regimes outperforms SR-od model in dynamical environment only if user embeddings are updated (Full and Rare regimes).
The right plot shows that regular updates of user embeddings improve performance.
    }
    \label{fig:ablation_study_app_usage}
\end{figure}
Figure~\ref{fig:ablation_study_app_usage} shows the performance of the SeqMF model in the dynamic environment for the App Usage dataset. 
From the left plot follows that the performance of the SeqMF model in Full and Rare regimes is very similar.
The possible reason for such a trend is that the first cycle contains enough information about patterns in user behavior.
Hence, it does not matter how often the users' embeddings are updated for this particular dataset.

\subsubsection{Dynamic environment: QHarmony mechanism outperforms competitors}
\label{subsec:privacy-study}
To analyze the impact of privacy mechanisms on the SeqMF model performance, we use a dynamic environment similar to the previous section and a Full training regime. 
We compare the performance of the SeqMF model without a privacy mechanism with the same model accompanied by the considered baselines and the proposed QHarmony mechanism.

\begin{figure}[h!]
\begin{subfigure}{.5\textwidth}
  \centering
  \includegraphics[width=.9\linewidth]{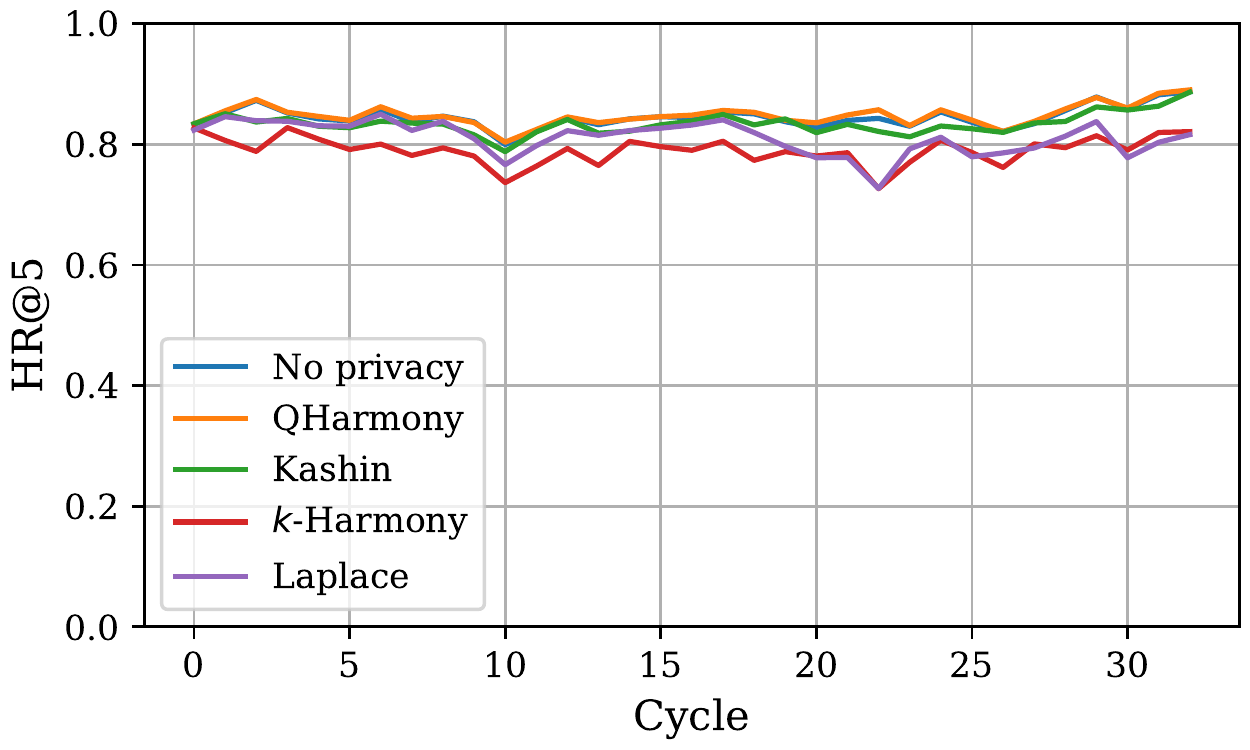}
\end{subfigure}%
\begin{subfigure}{.5\textwidth}
  \centering
  \includegraphics[width=.9\linewidth]{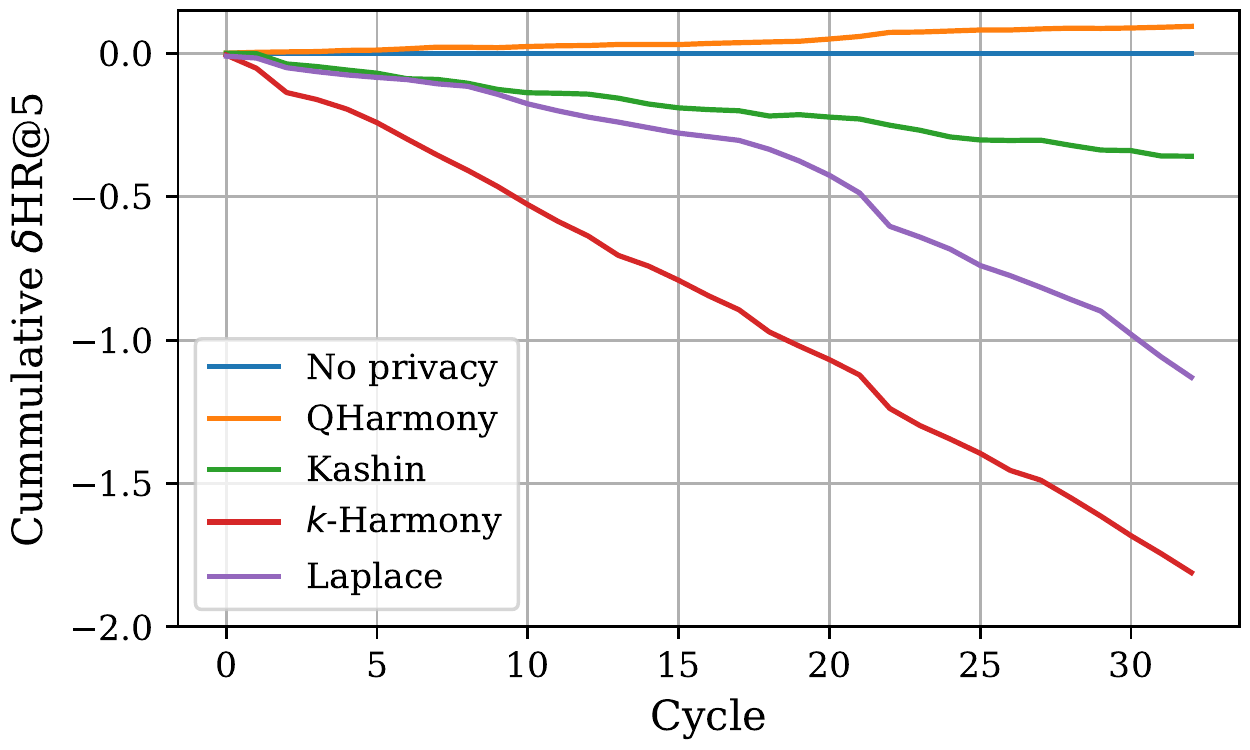}
\end{subfigure}
    \caption{\emph{LSApp dataset.} \textbf{Left:} Dependence of the HR@5 on the privacy-preserving mechanisms with privacy budget $\epsilon=4.5$ incorporated in the SeqMF model. 
    The higher HR@5 is, the better performance of the model is.
    Therefore, the QHarmony mechanism provides stronger protection for sensitive users' data. \textbf{Right:} The cumulative $\delta \mathrm{HR}@5$ presented for $33$ cycles confirms that the QHarmony mechanism does not lead to performance decreasing compared to the non-private SeqMF model.}
    \label{fig:privacy_study_lsapp}
\end{figure}

Figure~\ref{fig:privacy_study_lsapp} shows that the SeqMF model with the proposed QHarmony algorithm and privacy budget $\epsilon=4.5$ has an even higher utility on the LSApp dataset compared to the SeqMF model without any privacy mechanism. 
The possible reason for the such effect is the well-known implicit regularization property of stochastic gradient observed previously in deep learning studies~\cite{Neelakantan_noisy_grad_2015}.
At the same time, we observe the poor performance of the baseline privacy mechanisms except for the Kashin mechanism as was expected in Section~\ref{sec::q_harmony}.
Kashin mechanism still has the worse privacy-utility trade-off and is much more computationally costly than the proposed QHarmony mechanism.



\begin{figure}[h!]
\begin{subfigure}{.5\textwidth}
  \centering
  \includegraphics[width=.9\linewidth]{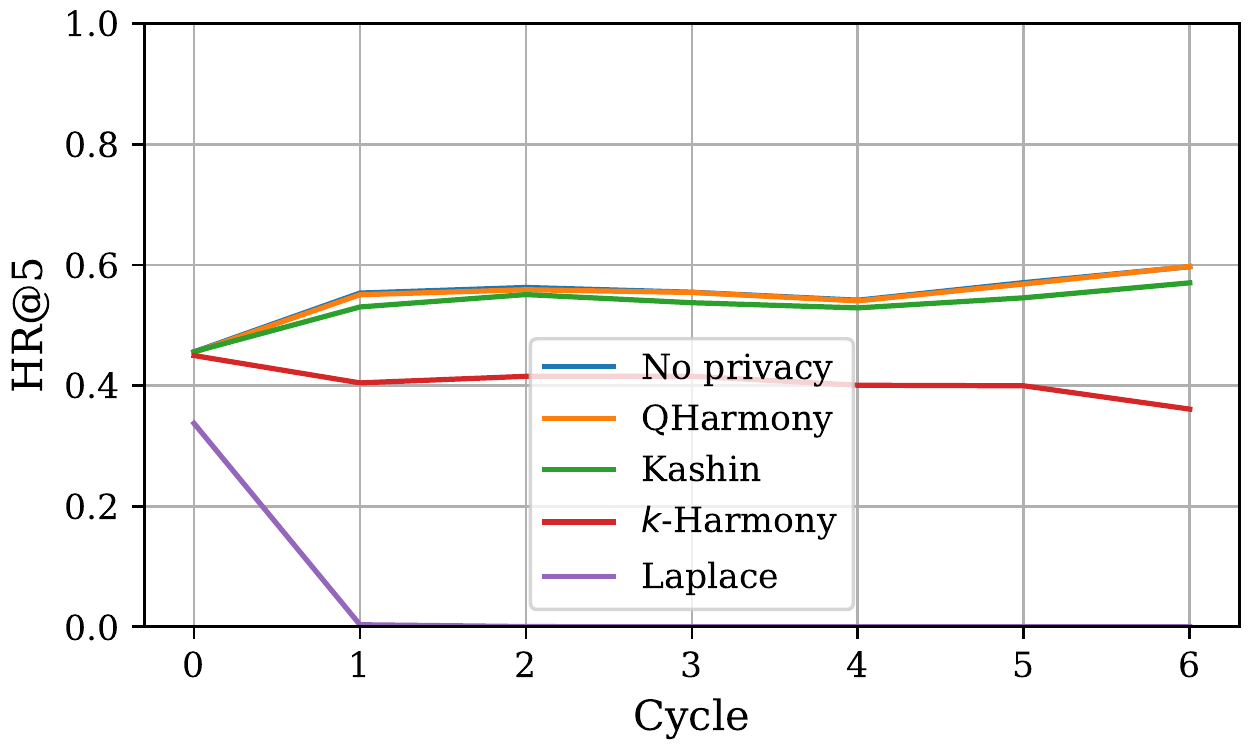}
\end{subfigure}%
\begin{subfigure}{.5\textwidth}
  \centering
  \includegraphics[width=.9\linewidth]{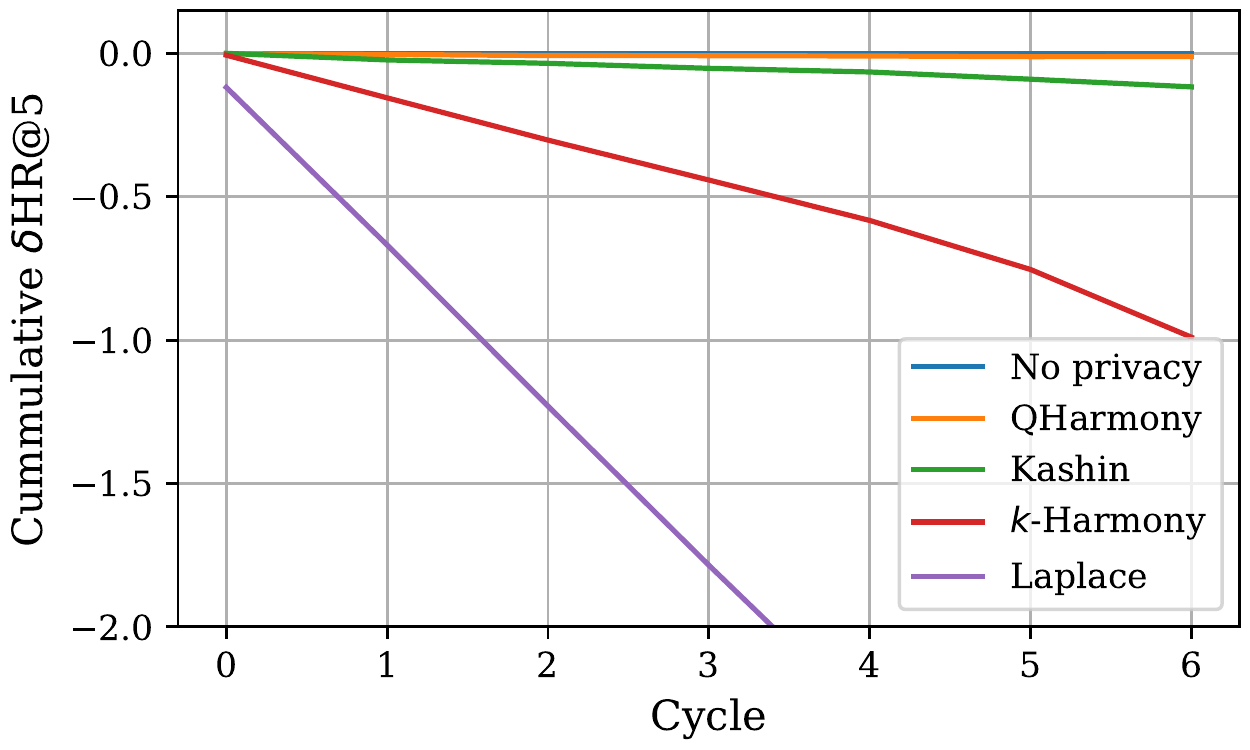}
\end{subfigure}
    \caption{\emph{App Usage dataset.} \textbf{Left:} Dependence of the HR@5 on privacy-preserving mechanisms with privacy budget $\epsilon=1.1$ incorporated to the SeqMF model. 
    The higher HR@5 is, the better performance of the model is. 
    The QHarmony mechanism provides stronger protection for sensitive users' data. \textbf{Right:} The cumulative $\delta \mathrm{HR}@5$ presented for $7$ cycles confirms that the QHarmony mechanism does not lead to performance decreasing compared to the non-private SeqMF model.}
    \label{fig:privacy_study_app_usage}
\end{figure}

Figure~\ref{fig:privacy_study_app_usage} shows that the proposed QHarmony mechanism with budget $\epsilon = 1.1$ preserved almost the same level of $\mathrm{HR}@5$ compared to the non-private SeqMF model on App Usage data. 
As in the case of the LSApp data, baseline privacy mechanisms have a worse privacy-utility trade-off compared to the proposed QHarmony mechanism. 
These experiments confirm the effectiveness of the proposed QHarmony mechanism and demonstrate that performance of the SeqMF model is preserved after incorporating the QHarmony mechanism for both considered datasets.

\section{Conclusion}
\label{sec:conclusion}
In this study, we considered the next-app prediction problem and proposed a new sequence-aware matrix factorization model called SeqMF.
The model can predict the next app for every user which can be used to optimize the resource consumption on the user's device.
SeqMF model takes into account the sequential nature of data, is accompanied by a new QHarmony mechanism for stronger privacy guarantees, and is trained in the federated learning paradigm.
This paradigm means that users' embeddings are stored locally on the user devices and apps' embeddings are the same for all devices and are updated in the remote server.
These features make the SeqMF model robust to potential leakage of the transmitted data from user devices to the remote server and help in capturing behavior patterns common for multiple users.
To illustrate the performance of our model we carried out experiments on the LSApp and App Usage datasets.
In these experiments, the real-world setting of data flow was emulated to evaluate considered models properly.
Our experiments demonstrate that the SeqMF model outperforms the standard matrix factorization model and other competitors based on the collected statistics.
Also, the proposed QHarmony privacy mechanism shows the best privacy-utility trade-off compared to the competitors and moderate computational complexity.
Thus, the proposed SeqMF model accompanied by the QHarmony privacy mechanism provides practical advantages in real-world applications over the previously proposed solutions for the next-app prediction task.



\bibliographystyle{abbrvnat}
\bibliography{main}






\end{document}